# MOF-derived Fe-doped $\delta$-MnO$_2$ nanoflowers as oxidase mimics: Chromogenic sensing of Hg(II) and hydroquinone in aqueous media

Udisha Duhan,[a] Arnab Pan,[b] Ritesh Dubey,[a] Samar Layek,*,[c] Sushil Kumar*,[a] and Tapas Goswami*,[a]

Structure and morphology play a crucial role in enhancing the biomimetic oxidase activity of nanozymes. In this study, a facile in-situ chemical oxidation strategy was employed to synthesize MOF-derived MnO$_x$, utilizing the structural features of the parent MOF to enhance oxidase-mimicking activity. We have systematically investigated the effects of phase evolution, structural modulation and morphology on the oxidase activity of MnO$_x$ with Fe substitution. The oxidase-like activity was evaluated using a chromogenic substrate 3,3′,5,5′-tetramethylbenzidine (TMB), which produced a blue colored ox-TMB with an absorption peak at 652 nm upon oxidation. While all Fe-doped MnO$_x$ nanostructures exhibited oxidase-like activity, the 10% Fe-doped sample (**10Fe-MnO$_x$**) demonstrated the highest performance, likely due to a synergistic effect of structure, morphology and existing oxygen vacancies. The underlying oxidase mechanism was investigated using steady-state kinetics and electron paramagnetic resonance (EPR) analysis. In addition, a colorimetric assay was developed for detecting Hg$^{2+}$ and hydroquinone (HQ) in real water samples collected from industrial and natural sources. The calculated detection limits of the colorimetric probe **10Fe-MnO$_x$** for HQ (1.74 μM) and Hg$^{2+}$ (0.47 μM) outperformed conventional metal oxide-based nanozymes. These findings may pave a way for the development of easily synthesizable, scalable and highly sensitive oxidase-based MOF-derived metal oxide nanomaterials having applied potential in biological and environmental settings.

## Introduction

Over the preceding years, natural enzymes have been the subject of extensive research in various biochemical processes due to their high substrate specificity and catalytic efficacy.[1-3] However, their catalytic activity is susceptible to several factors such as pH, temperature, reaction medium, etc. Furthermore, their widespread application is limited by high production costs, storage challenges, instability (including denaturation and protease digestion) and purification difficulties.[3,4] Significant efforts are thus underway to develop novel enzyme mimics exhibiting desired catalytic activity and chemical stability. The artificial enzymes offer several advantages over their natural counterparts, including enhanced stability, greater tolerance to environmental variations, and low-cost production and storage.[5-8] Contextually, a diverse range of synthetic materials have been documented to exhibit bio-mimetic activity similar to natural enzymes such as peroxidases,[9] oxidases,[10] catalases,[11] superoxide dismutase,[12] etc. To develop artificial enzymes, inorganic materials have emerged as more promising synthetic models due to their tuneable structural, electronic and redox features.[13] Inspired from the work by Berfield et al. which demonstrated that Fe$_3$O$_4$ can catalyse the oxidation of chromogenic substrate tetramethylbenzidine (TMB),[14] recent studies have explored the enzymatic potential of various nanomaterials based on CuO,[15] Co$_3$O$_4$,[16] V$_2$O$_5$,[17] MnO$_2$,[18] Ag$_2$O,[19] etc. Among them, manganese oxides (MnO$_x$) have gained particular attention due to a variable redox behaviour, cost-effectiveness, favourable biological safety profile and relatively higher oxidizing activity.[20] MnO$_x$ can be developed in multiple phases (such as MnO, Mn$_3$O$_4$, Mn$_2$O$_3$ and MnO$_2$), with MnO$_2$ standing out for its superior catalytic efficacy and recyclability. Moreover, MnO$_2$ nanostructures can be easily tuned to attain a desired morphology including nanotubes, nanowires, nanoplates and nanorods.[21] For instance, Zhang et al. synthesized MnO$_2$ nanowires, MnOOH/MnO$_2$ nanorods, and octahedral MnO$_2$ nanoparticles using a hydrothermal method, finding that MnO$_2$ nanowires exhibited superior catalytic activity as oxidase mimics.[22] Similarly, Liu et al. demonstrated the oxidase-mimicking activity of MnO$_2$ nanoparticles, showcasing their ability to catalyse the oxidation of colorless TMB to blue oxidized TMB (ox-TMB) without the need for H$_2$O$_2$.[23] Despite these advancements, studies on the structure-activity relationship of MnO$_2$ oxidase mimics remain limited. Structurally, MnO$_2$ exists in various crystallized form (such as $\alpha$-, $\beta$- and $\delta$-MnO$_2$) each defined by its number of [MnO$_6$] octahedral subunits framework and tunnel size.[24,25] Furthermore, doping of various metal ions such as V$^{5+}$, Ag$^+$, Co$^{2+}$, Ni$^{2+}$, Cu$^{2+}$ and Fe$^{+3/+4}$ in MnO$_2$ substantially improve the redox properties, dispersion of active sites which are crucial for catalytic applications.[26-32] Among these, iron (Fe) is particularly beneficial as it can either substitute Mn or occupy the tunnel structures, stabilizing the material and improving catalytic efficiency.[33] Wang et al. reported that Fe doping influences the phase of MnO$_2$ nanostructures, with different Fe concentrations yielding $\alpha$-MnO$_2$, R-MnO$_2$, or $\beta$-MnO$_2$, thereby enhancing capacitance retention and energy density.[34] Moreover, Fe doping introduces oxygen vacancies, which are known to significantly enhance electrocatalytic and photocatalytic activities due to the increased active sites and high surface

[a.] Department of Chemistry, Applied Sciences Cluster, UPES Dehradun, Energy Acres Building, Dehradun- 248007, Uttarakhand, India.
[b.] Light Stock Processing Division, CSIR – Indian Institute of Petroleum, Dehradun- 248005, Uttarakhand, India.
[c.] Department of Physics, Applied Sciences Cluster, UPES Dehradun, Energy Acres Building, Dehradun- 248007, Uttarakhand, India.

energy.[35] While these reported nanozymes exhibit comparable catalytic activities to natural enzymes, their intrinsic catalytic activities are often compromised by factors such as a limited number of exposed active sites, intrinsically deficient multilevel structures as well as aggregation issues.[36-38]

Among them, $MnO_2$ nanostructures with high porosity and large number of exposed active sites are found more promising for a specific catalytic application. In this endeavour, $MnO_x$ derived from metal organic frameworks (MOFs) have emerged as a profound class of porous materials.[39-41] In particular, MOF-derived $MnO_2$ nanostructures have attracted increasing research interest in catalysis due to their well-defined coordination networks and tunable porosity. The diversity of metallic nodes, linking struts, and coordination interactions in all possible directions position MOFs as direct surrogates to develop metal oxides based artificial enzymes. The well-defined, tailorable cavities and channels of parent MOFs can be delivered to the resulting metal oxides. Typically, the rationale behind the use of MOF-derived metal oxides as promising catalysts lies in their ability to retain high surface area, enhancing catalytic activity by exposing more active sites. The MOF-to-MO conversion also enables precise control over crystal phase, particle size, and morphology. Carbon residues from the MOF structure provide stability to MOs by encapsulating the nanostructures and preventing active component loss during reactions. The porous architecture ensures efficient mass transfer and improved interaction between catalysts and reactants. Based on these advantages, several MOFs have shown potential as templates, suitable support materials, and precursors for manufacturing nanostructured materials. Compared to the existing mesoporous materials, MOFs derivatives offer a novel pathway for constructing mesoporous nanozymes. In this context, the intrinsic mesoporous properties of MOFs can be maintained, while the stability and activity of the resulting nanozymes can be significantly improved. Furthermore, the mesoporous structures and multiple channels can enable full contact with substrates and facilitate mass transport and diffusion of products in catalytic applications.[42-45]

Recently, several studies have demonstrated that MOFs-derived manganese oxides offer precise control over structure and significantly improve their catalytic properties. For example, Lu et al. synthesized a hollow MnFeO oxide from a PBA@PBA core-shell structure via thermal treatment, achieving both oxidase- and peroxidase-like activities.[46] Similarly, $Fe_3O_4$@C@$MnO_2$ composites, formed via an interfacial reaction between pyrolyzed Fe-MIL-88A and $KMnO_4$, exhibited appreciable biomimetic activities.[38] Gai et al. developed $MnO_x$@NC with dual-enzyme properties by calcining a Mn-TTPCA MOF.[42] While these studies highlight the potential of MOF-derived $MnO_2$ nanozymes, their thermal synthesis often results in aggregation or structural collapse, limiting their catalytic efficiency. A more controlled in-situ chemical oxidation strategy is thus essential to optimize the structure and morphology of such materials. To the best of our knowledge, the impact of Fe doping on the phase, structure, and morphology of $MnO_x$ nanostructures and its influence on oxidase activity remains largely unexplored.

Herein, we introduce a new cost-effective strategy to prepare MOF-derived Fe-doped $\delta$-$MnO_2$ nanoflowers via straightforward solvent treatment of MOF template. The as prepared $MnO_x$ successfully retained the morphology of the parent MOF while exhibiting phase transitions influenced by varying levels of Fe doping. Notably, at 10% Fe doping, a complete conversion to a pure $\delta$-$MnO_2$ phase was observed. The oxidase activities of MOF-derived manganese oxides were assessed by monitoring the oxidation of TMB in absence of hydrogen peroxide ($H_2O_2$). All synthesized materials demonstrated the oxidase activity; however, the 10% Fe-doped $MnO_x$ (referred as **10Fe-$MnO_x$**) exhibited the highest catalytic activity, attributed to its phase structure, nanoflower-like morphology and oxygen vacancies. The catalytic mechanism was investigated using steady-state kinetics and EPR study. Furthermore, under optimized conditions, we demonstrated a straightforward colorimetric method for sensitive detection of hydroquinone and $Hg^{2+}$ ion.

## Experimental Section

**Preparation of MnBTC MOF and Fe-doped MnBTC MOFs.** MnBTC MOF was prepared using a procedure reported elsewhere.[47] Briefly, 2.0 g PVP ($M_w$ = 30,000) was dissolved in a 100 mL solvent mixture of ethanol and water (2:1, v/v) and kept on stirring for 10 mins at 400 rpm. Thereafter, 2.0 mmol of Mn($CH_3COO$)$_2$ was added to the above solution to form a colourless transparent solution A. In a separate beaker, 4.73 mmol (0.9 g) of 1,3,5-benzenetricarboxylic acid (BTC) was dissolved in 100 mL ethanol-water (2:1, v/v) mixture to obtain colorless solution B. The two solutions A and B were then mixed and stirred for 1 hour at 400 rpm, and the resulting mixture was left for rest overnight at room temperature. The white-colored powder was collected by centrifugation, washed thrice with ethanol/water and finally dried at 60 °C for 4 hours.

A series of Fe-doped MnBTC MOFs were synthesized following the similar procedure described above, with varying Fe molar percentages (1%, 5% and 10%) relative to Mn while keeping the total metal precursors amount (Mn($CH_3COO$)$_2$ and $FeCl_3$) at 2 mmol. The resulting products were labelled as 1Fe-MnBTC, 5Fe-MnBTC and 10Fe-MnBTC, respectively.

**Preparation of MOF derived $MnO_x$ and Fe-doped $MnO_x$.** MOF-derived manganese oxides were synthesized using a facile chemical oxidation method. An alkaline NaOCl solution was prepared by mixing 100 mL of a 0.5 M NaOH aqueous solution with 100 mL of a 0.3 mmol NaOCl aqueous solution at room temperature. Subsequently, 0.05 g of the synthesized MnBTC MOF was added to the alkaline solution of NaOCl and stirred overnight to obtain a black suspension of $MnO_x$. The resulting suspension was centrifuged, washed three times with ethanol/water, and dried under vacuum to yield black-colored powder of $MnO_x$.

Fe-doped $MnO_x$ were prepared using the same chemical oxidation method applied to 1Fe-MnBTC, 5Fe-MnBTC and 10Fe-MnBTC, resulting in 1Fe-$MnO_x$, 5Fe-$MnO_x$, and 10Fe-$MnO_x$, respectively.

## Results and discussion

**Structural description.** Fe-doped $MnO_x$ were synthesized through a simple chemical oxidation of Fe-MnBTC MOFs under alkaline conditions using NaOCl. Weak Mn(II)-O bonds in Mn-BTC MOFs facilitate their easy transformation into metal oxides *via* solvent treatment. Upon exposure to NaOH, OH⁻ ions diffuse into the porous MOF structure, displacing BTC ligands through a thermodynamically favorable process and forming an unstable $Mn(OH)_2$ intermediate. This intermediate subsequently reacts with ClO⁻ ions in solution, resulting in the formation of a mixed-phase $MnO_x$. The reaction process can be represented as follows:

$Mn_3(BTC) + 6OH^- \rightarrow 3Mn(OH)_2 + BTC^{3-}$
$Mn(OH)_2 + ClO^- \rightarrow Cl^- + MnO_2(s) + H_2O$

The phase structure of the parent Fe-MnBTC MOFs was first examined using PXRD analysis, as shown in Fig. S1. The PXRD patterns closely matched with the reported and simulated patterns, confirming the successful synthesis of the MOFs. With increasing Fe doping, a slight peak shift toward higher angle was observed, suggesting minimal structural alterations. This shift suggests a gradual substitution of Mn by Fe while preserving the original crystal structure. To investigate the structural and phase evolution with varying Fe-doping levels, the PXRD patterns of MOF-derived manganese oxides were analyzed (Fig. 1). The undoped MnBTC MOF-derived manganese oxide exhibited a mixed-phase composition, primarily consisting of the tetragonal $Mn_3O_4$ phase (JCPDS: 24-0734) and the monoclinic *δ*-$MnO_2$ phase (JCPDS: 80-1098). The weak diffraction peaks indicated a predominantly amorphous nature, while the broad peaks suggested a nanoscale morphological structure. Although the phase composition of $MnO_x$ remained largely unchanged upon 1% and 5% Fe doping, a significant improvement could be realized in the crystallinity as evidenced by enhanced peak intensities. The resulting higher crystallinity indicates that the incorporation of small amounts of Fe may promote better ordering within the crystal structure without altering the crystal phases.

A drastic transformation occurred at 10% Fe doping, where the phase composition shifted to pure *δ*-$MnO_2$. The PXRD pattern of **10Fe-MnO$_x$** displayed distinct peaks dedicated solely to *δ*-$MnO_2$, confirming a complete transition from a mixed-phase $MnO_x$ to *δ*-$MnO_2$ at this doping concentration. This phase change suggests that 10% Fe doping disrupts the stability of $Mn_3O_4$ under oxidizing conditions, favouring the formation of pure *δ*-$MnO_2$. In this process, Fe acted as both a structure-directing agent and a catalytic promoter, enabling the oxidation of $Mn_3O_4$ to $MnO_2$.

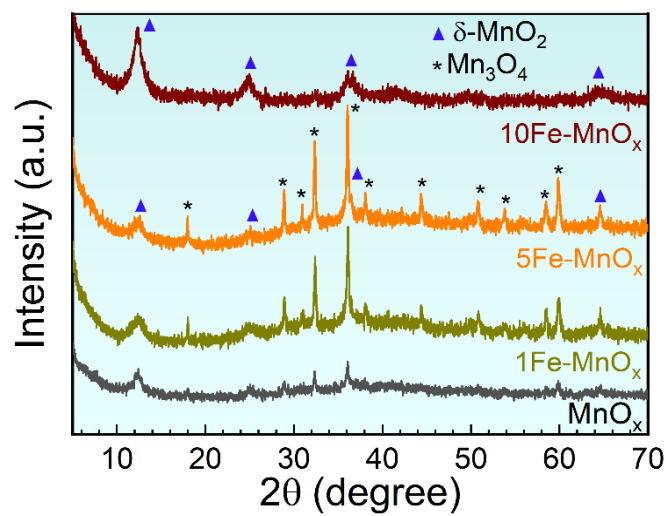

**Fig. 1** PXRD spectra of MOF-derived MnO$_x$, **1Fe-MnO$_x$**, **5Fe-MnO$_x$** and **10Fe-MnO$_x$**.

While several research groups have attributed sharp peaks at similar 2θ values to the *α*-$MnO_2$ phase having orthorhombic crystal structure, a few reports suggest $Mn_3O_4$ phase with tetragonal structure. We resolved this ambiguity using Rietveld refinement analysis, confirming the peaks correspond to $Mn_3O_4$. Rietveld refinement analysis was performed on PXRD data of **5Fe-MnO$_x$** and **10Fe-MnO$_x$** nanostructures using the GSAS-II software[48] to confirm the initially identified phases and quantify their percentages (Fig. 2). Table S1 presents the lattice parameters and crystallographic angles derived from the best-fit data, along with relevant fitting parameters. The XRD pattern for the **5Fe-MnO$_x$** nanostructures was best modelled as a mixture of tetragonal $Mn_3O_4$ (78%, space group $I4_1/amd$) and monoclinic *δ*-$MnO_2$ (22%, space group $C2/m$). In contrast, the **10Fe-MnO$_x$** nanostructures were found to be purely *δ*-$MnO_2$ with a monoclinic crystal structure (space group $C2/m$). The Rietveld analysis also confirm that in both the sample Fe-occupies some of the Mn site in manganese oxide crystal lattice. These analyses provide insights into the influence of iron doping concentration on phase formation, crystal structures, and microstructures, which will be discussed in subsequent sections.

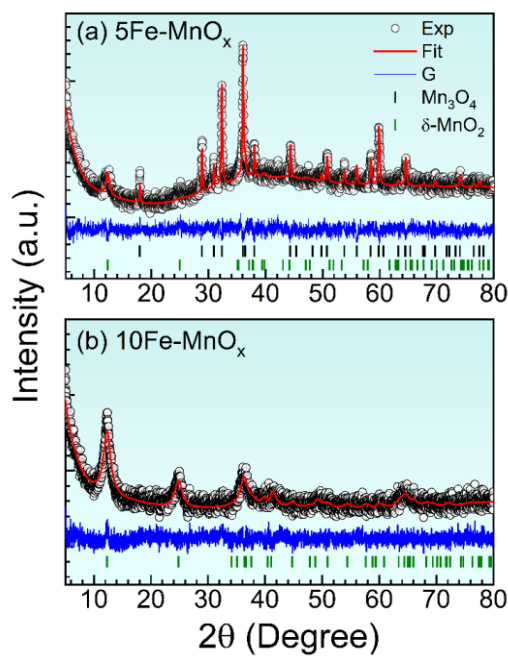

**Fig. 2** Rietveld refinement plots of XRD patterns for (a) **5Fe-MnO$_x$** and (b) **10Fe-MnO$_x$** nanostructures. The experimental data is shown as open black circles, the theoretical fit as solid red lines, and the Bragg positions for Mn$_3$O$_4$ and $\delta$-MnO$_2$ crystal structures are indicated by black and olive bars, respectively. Blue solid lines at the bottom reflects the difference between experimental and theoretical data.

The FE-SEM and TEM analyses have been conducted to investigate the morphology and microstructures of the as-prepared Fe-MnBTC MOFs and **Fe-MnO$_x$**. The SEM images of Mn-MOF revealed the presence of urchin/flower-like architectures composed of thin nanosheets (Fig. 3a). FE-SEM micrographs of MnO$_x$ (Fig. 3b) displayed that the flower like morphology of the parent MOF is well preserved in the respective MnO$_x$.

Further comparison of SEM images of the MnO$_x$ samples revealed the impact of Fe doping on morphology. The undoped MnO$_x$ exhibited a mixed phase of Mn$_3$O$_4$ and $\delta$-MnO$_2$, characterized by an amorphous morphology resembling flower-like structures. In contrast, a notable transformation in morphology was observed at 5% Fe doping. The sample displayed a mixed morphology comprising both rod-like and flower-like structures (Fig. 3c). These results corroborate with the PXRD analysis exhibiting the improved crystallinity in **5Fe-MnO$_x$**. At 10% Fe-doping, the morphology shifted largely to flower-like structures (Fig. 3d). This transition indicates that the higher levels of Fe-doping favour the establishment of a more uniform flower-like morphology. It is worth mentioning that the incorporation of Fe into framework likely influences the nucleation and growth trend during synthesis of MOF derived metal oxide. 10% Fe$^{3+}$ dopants can significantly alter the coordination environment and surface energy of growing crystallites, thus promoting a more controlled and homogeneous self-assembly process. The results from the FE-SEM analysis were well aligned with those obtained from PXRD analysis, highlighting that variation in Fe levels not only influences the phase composition and crystallinity but also significantly affects the morphological characteristics of MnO$_x$.

Furthermore, the morphology, microstructure and composition of the as-prepared **5Fe-MnO$_x$** and **10Fe-MnO$_x$** were evaluated using TEM analysis (Fig. 4). The TEM images of **5Fe-MnO$_x$** (Fig. 4a) revealed the presence of nanorods like morphology. This observation was consistent with the SEM analysis. High-resolution TEM (HR-TEM) micrographs showed a lattice fringe spacing (d) of ~0.49 nm, corresponding to the (101) planes of the tetragonal Mn$_3$O$_4$ structure (Fig. 4b). Additionally, a flower-like morphology was observed, featuring stripe spacing of ~0.592 nm, a characteristic fingerprint of birnessite-type $\delta$-MnO$_2$. The SAED image in Fig. 4c indicated polycrystallinity and the rings are in good agreement with mixed phase pattern of Mn$_3$O$_4$ and $\delta$-MnO$_2$. Similar to the SEM analysis, a complete transfer of morphology to nanoflowers could be seen when 10% Fe was doped (Fig. 4d). These nanoflowers exhibited stripe spacings of ~0.72 nm, attributed to the presence of interlayer water molecules and cations that stabilize the layered sheets. This d-spacing corresponds to the (001) reflection planes of $\delta$-MnO$_2$ (Fig. 4e), which could also be observed in the PXRD pattern at 2θ = 12.6°. The layer thickness of **10Fe-MnO$_x$** was determined to be ~ 3.9 nm, which is consistent with the crystallite size measured from PXRD pattern (Fig. S2). This atomic-level thickness and layered morphology likely enhance the exposure of active sites, facilitating electron transfer and thereby contributing to the superior catalytic activity. The SAED pattern of **10Fe-MnO$_2$** exhibited small spots along the rings, indicating its polycrystalline nature (Fig. 4f). All the rings closely matched the d-values associated with $\delta$-MnO$_2$. This structural evolution highlights the significant impact of Fe doping on the morphology and crystal structure of MnO$_x$, promoting the transition from a mixed morphology to a nanoflower-like structure with enhanced layer stability. Furthermore, the EDS elemental mapping of both **5Fe-MnO$_x$** and **10Fe-MnO$_x$** demonstrated a homogenous distribution of elements throughout the nanostructure, with C, O, Mn and Fe as the main elements (Fig. S3 and Fig. S4). The EDX spectra of **5Fe-MnO$_x$** and **10Fe-MnO$_x$** confirmed %Fe doping as 6 and 9.4, as shown in the inset table of Fig. S5.

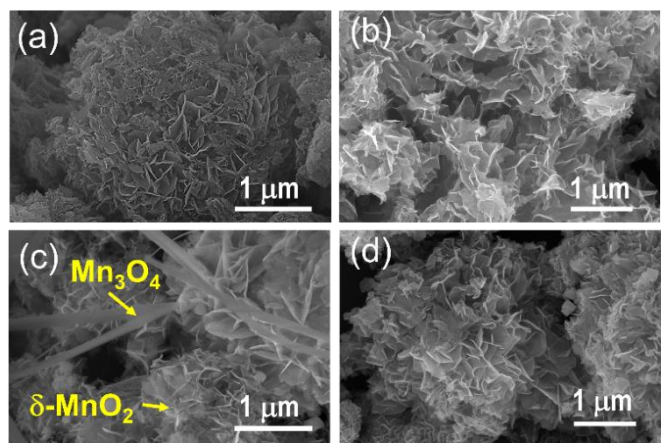

Fig. 3 FE-SEM micrographs of (a) Mn-MOF; (b) MnO$_x$; (c) **5Fe-MnO$_x$** and (d) **10Fe-MnO$_x$**.

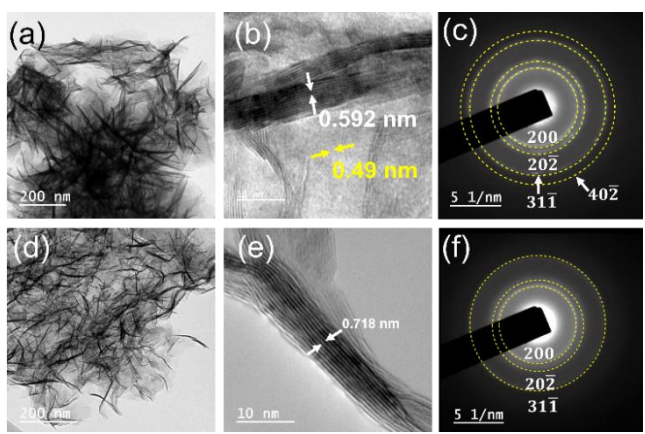

**Fig. 4** High resolution TEM images and SAED patterns of **5Fe-MnO$_x$** (a-c) and **10Fe-MnO$_x$** (d-f).

To analyze the chemical state and composition of the MOFs and their derived manganese oxides, X-ray photoelectron spectroscopy (XPS) was conducted. The XPS survey spectra (Fig. S6) confirmed the presence of Mn, O and Fe, with no detectable impurity peaks. The Mn 2p spectra exhibited two main peaks at ~642.6 eV and ~654.0 eV, corresponding to Mn 2p$_{3/2}$ and Mn 2p$_{1/2}$, respectively. High-resolution spectra were deconvoluted using XPSPEAK41 software. For all manganese oxide samples, deconvolution of the Mn 2p$_{3/2}$ peak revealed three distinct oxidation states: Mn$^{4+}$ (~644.1 eV), Mn$^{3+}$ (~642.7 eV) and Mn$^{2+}$ (~641.1 eV) (Fig. 5). However, the relative contributions of Mn$^{2+}$, Mn$^{3+}$ and Mn$^{4+}$ varied significantly among samples. Notably, the **10Fe-MnO$_x$** sample exhibited a higher surface concentration of Mn$^{3+}$ and Mn$^{4+}$ compared to the 0% and 5% Fe-doped samples. Additionally, shifts in the 2p$_{3/2}$ and 2p$_{1/2}$ peak positions of Mn could be noticed at higher binding energies with increasing Fe-doping, along with changes in the Mn$^{2+}$/Mn$^{3+}$/Mn$^{4+}$ ratios, indicated the modulation of oxidation states along with related spin-orbit coupling upon Fe incorporation. The deconvolution of the asymmetric O 1s spectra for all MOF-derived manganese oxides revealed three distinct components with binding energies approximately at 529.6 eV, 530.7 eV and 532.8 eV (Fig. 5). Based on the existing literature, these peaks correspond to lattice oxygen (O$_{latt}$, Mn-O-Mn), oxygen vacancies (O$_V$), and surface-adsorbed oxygen (O$_{ads}$, Mn-O-H), respectively. The oxygen vacancy concentration was quantified using the ratio [O$_V$/(O$_L$+O$_V$+O$_{ads}$)]. The Fe doping level significantly influenced oxygen vacancy concentrations, with the **10Fe-MnO$_x$** sample exhibiting the highest value (19%), followed by **5Fe-MnO$_x$** (15%) and **0Fe-MnO$_x$** (12%). An increased oxygen vacancy concentration correlates with a greater number of active oxygen sites, which can eventually enhance catalytic oxidation performance.

High-resolution Fe 2p spectral analysis for the **5Fe-MnO$_x$** and **10Fe-MnO$_x$** samples is presented in Fig. S7. In both samples, the Fe spin-orbit peaks 2p$_{1/2}$ and 2p$_{3/2}$ were observed at binding energies of 724.7 eV and 710.6 eV, respectively. The deconvolution of these spin-orbit peaks revealed three distinct peaks: two at approximately 710.2 eV and 711.5 eV, corresponding to Fe$^{3+}$, and a smaller peak at 713.5 eV, attributed to Fe$^{2+}$. This indicates the presence of both Fe$^{3+}$ and Fe$^{2+}$ in the MnO$_x$ lattice, with varying Fe doping levels. Notably, the intensity of the Fe 2p peaks increases with higher Fe doping. Additionally, satellite peaks have been observed at higher binding energies, specifically at 719.3 eV and 732.6 eV. The synthesized **10Fe-MnO$_x$** shows promising potential for oxidase-mimic catalysis, attributed to the solid-state redox couples of Mn$^{3+}$/Mn$^{4+}$, Fe$^{3+}$/Fe$^{2+}$ and the presence of oxygen vacancies.

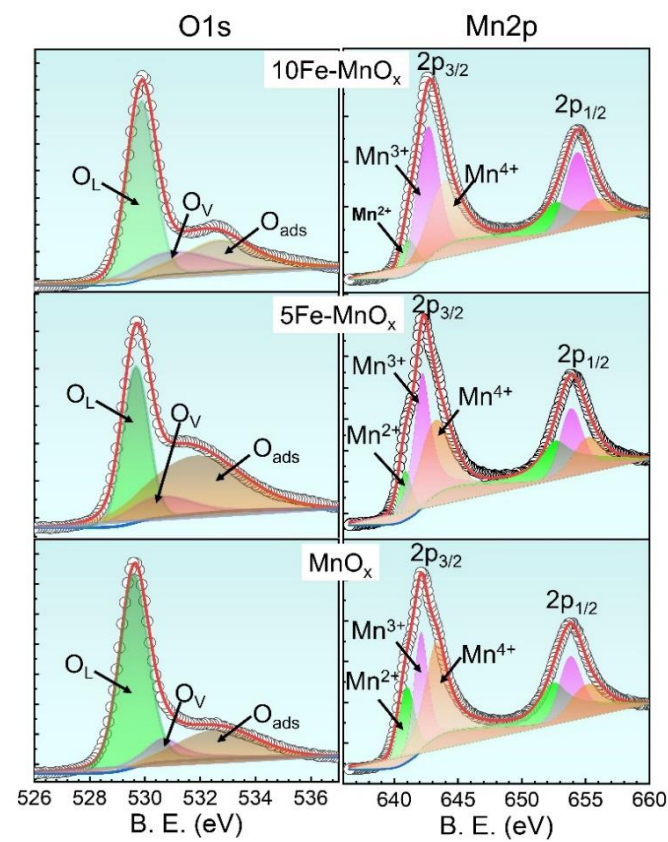

**Fig. 5** Deconvoluted XPS spectra of O1s and Mn 2p for MnO$_x$, **5Fe-MnO$_x$** and **10Fe-MnO$_x$**.

**Oxidase-like activity of manganese oxides.** The synthesized manganese oxides have been employed as catalysts to explore their potential in biomimetic oxidase-like activity. It is well known that the reaction conditions and environmental factors significantly influence the oxidase-like activities of both natural and artificial enzymes. In order to evaluate the impact of Fe doping in MnO$_x$, key catalytic parameters including incubation time, pH, temperature, and catalytic dosage of **10Fe-MnO$_x$** were optimized for the oxidation of a chromogenic substrate, TMB. UV-Vis study was conducted to monitor the activity of **10Fe-MnO$_x$** for a period of 20 mins where the absorbance at 652 nm rapidly increased and reached saturation within 15 mins (Fig. S8a). Although **10Fe-MnO$_x$** exhibited the highest activity at 15 °C, most likely due to the increased solubility of dissolved oxygen at lower temperatures, all subsequent experiments were conducted at room temperature for practical relevance (Fig. S8b). pH optimization is also crucial in enzyme-based catalysis, as biological enzymes often exhibit varying activity across different pH levels. The catalytic activity of **10Fe-MnO$_x$** was thus tested in different buffer solutions with pH ranging from 2.0 to 10.0. As illustrated in Fig. S8c, the oxidase activity of **10Fe-MnO$_x$** increased between pH 2.0 to 4.0, reaching the highest response at pH 4.0. Further increase in pH led to a decline in catalytic efficiency, and therefore, pH 4.0 was selected as the optimal value. Additionally, to determine the ideal catalyst loading, various concentrations of **10Fe-MnO$_x$** were evaluated for TMB oxidation. As shown in Fig. S8d, the highest catalytic efficiency was achieved at an optimal dosage of 6 μg/mL.

Under optimized conditions, the oxidase-like activity of all MOF-derived MnO$_x$ samples was systematically compared. A notable trend emerged, wherein progressive Fe-doping led to a steady enhancement in catalytic performance, with **10Fe-MnO$_x$** exhibiting the highest activity. This was further validated through UV-Vis absorption studies. To quantitatively assess the oxidase-like efficiency of both undoped and Fe-doped MnO$_x$, the absorbance change of TMB at 652 nm was monitored. Among all samples, **10Fe-MnO$_x$** demonstrated the most pronounced absorption peak at 652 nm, confirming its superior catalytic activity (Fig. 6a).

The relative activity of different catalysts was calculated using the formula as described in experimental section. The catalytic efficiency of the MOF-derived MnO$_x$ samples followed the order: MnO$_x$ < **1Fe-MnO$_x$** < **5Fe-MnO$_x$** < **10Fe-MnO$_x$** (Fig. 6b). The exceptional performance of **10Fe-MnO$_x$** can be attributed to its unique phase structure and nanoflower-like morphology, which features interlayer channels and cavities facilitating superior substrate diffusion, thereby enhancing catalytic activity. Fe$^{3+}$ doping into the MnO$_x$ lattice, which primarily contains Mn$^{4+}$, induced the formation of oxygen vacancies and promoted the presence of redox-active Mn$^{3+}$/Mn$^{4+}$ and Fe$^{3+}$/Fe$^{2+}$ couples- both of which are critical for oxidation reactions. XPS analysis further confirmed that **10Fe-MnO$_x$** exhibited the highest concentration of oxygen vacancies among all tested catalysts, reinforcing its superior catalytic efficiency.

For comparison, the oxidase-like activity of the parent MOF and **10Fe-MnO$_x$** synthesized via a co-precipitation method was evaluated. The detailed synthesis procedure is provided in the ESI, and the PXRD pattern (Fig. S9) of the 10% Fe-doped MnO$_x$ confirms the formation of $\delta$-MnO$_2$. As shown in Fig. S10, the MOF-derived **10Fe-MnO$_x$** displays significantly enhanced activity, highlighting the advantage of the MOF-templated approach in generating catalytically active Fe-doped $\delta$-MnO$_2$.

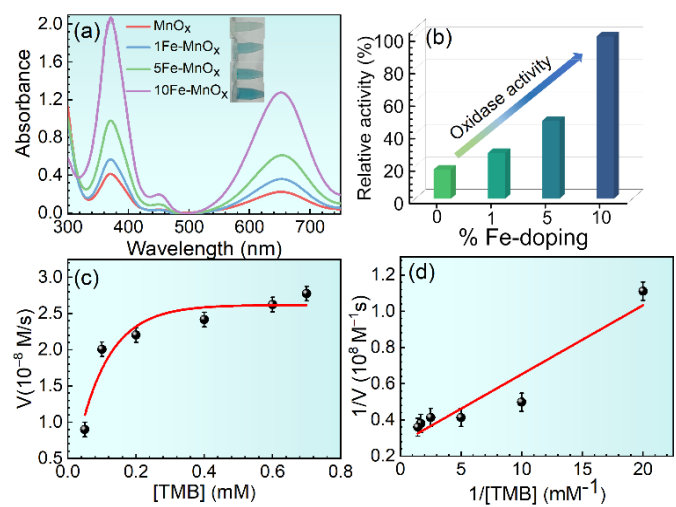

**Fig. 6** (a-b) Comparative performance of different MOF-derived MnO$_x$ catalysts in oxidase-like activity (assay conditions: catalyst loading 6.6 μg/mL; pH = 4.0; r. t.; TMB = 0.1 mM). Inset: photograph of the TMB-buffer system with different catalysts. Steady-state kinetic analysis using (a) Michaelis-Menten curves and (b) Lineweaver-Burk plots for catalytic oxidation of TMB using **10Fe-MnO$_x$**.

**Kinetic studies of oxidase mimicking activity.** A steady-state kinetic analysis was conducted under optimized conditions by varying TMB concentrations (0-100 μM) while maintaining all other parameters constant. The formation rate of the oxidized product was monitored by tracking the absorbance increase at 652 nm. Using the molar extinction coefficient ($\varepsilon$ = 3900 M$^{-1}$cm$^{-1}$), the corresponding concentrations of ox-TMB were calculated. Michaelis-Menten curves presented in Fig. 6c shows the relationship between V$_0$ and TMB concentration. The Lineweaver-Burk plots (Fig. 6d) were obtained by plotting the double reciprocal 1/V vs 1/[TMB], in which the slope of the double reciprocal curve corresponds to the K$_m$/V$_{max}$ ratio, and the intercept of the double reciprocal curve corresponds to 1/V$_{max}$. The K$_m$ value represents the affinity between the substrate TMB and the catalyst, where a lower K$_m$ value often indicates a strong affinity of substrate to the catalyst.[49] As summarised in Table S2, compared to HRP and other reported bioinspired enzymes, **10Fe-MnO$_x$** exhibited a relatively lower K$_m$ value (0.15 mM), indicating its higher catalytic activity. The superior activity of **10Fe-MnO$_x$** was mainly due to the large number of catalytically active sites generated via oxygen vacancies. Meanwhile, the good dispersive effect of MOF-derived metal oxides avoids the shortcoming of easy agglomeration, thus increasing the effective substrate-catalyst interactions.

**Oxidase mechanism.** Synthetic enzyme mimics typically generate ROS, which play a critical role in the oxidation of chromogenic substrates such as TMB. The oxidase-like activity of these enzymes relies on dissolved oxygen, which acts as an electron acceptor to generate superoxide radical anion ($O_2^{\cdot-}$). These active species can spontaneously convert into hydrogen peroxide ($H_2O_2$), which further decomposes to produce hydroxyl radicals (•OH) or singlet oxygen ($^1O_2$). As previously discussed, **10Fe-MnO$_x$** contains a high density of oxygen vacancies and redox-active metal couples ($Fe^{3+}/Fe^{2+}$, $Mn^{3+}/Mn^{4+}$), both of which facilitate efficient ROS generation. These ROS oxidize TMB into its blue-colored oxidized form, confirming the oxidase-mimicking activity of **10Fe-MnO$_x$**. To verify the *in-situ* generation of ROS, we employed DPPH (2,2-diphenyl-1-picrylhydrazyl), a stable free radical, as a detection probe. The presence of ROS was confirmed by a decrease in absorbance of DPPH near 520 nm, indicating ROS-mediated radical scavenging (Fig. S11). To specifically confirm the role of $O_2^{\cdot-}$ in TMB oxidation, we performed the reaction in the presence of *p*-benzoquinone, a known radical scavenger for $O_2^{\cdot-}$ radical anion. As shown in Fig. 7a, the relative oxidase-like activity of **10Fe-MnO$_x$** gradually decreased with increasing concentrations of *p*-benzoquinone, confirming the *in-situ* generation of $O_2^{\cdot-}$ to catalyze TMB oxidation.

To further confirm the ROS formation during oxidase-like activity, EPR studies were conducted with and without **10Fe-MnO$_x$** in ethanol (Fig. 7b). A strong hyperfine signal was observed in the EPR spectra within the 280–360 gauss range when 5.0 mg of **10Fe-MnO$_x$** was dispersed in ethanol. In contrast, no signal was detected in the control experiment (absence of the catalyst), confirming the formation of ROS in the colloidal suspension of **10Fe-MnO$_x$**. Additionally, the oxidation of TMB under aerobic and anaerobic conditions was examined to assess the role of dissolved oxygen or surface hydroxyls in the oxidase-like activity of **10Fe-MnO$_x$**. A significant reduction in absorbance at 652 nm was observed when $N_2$ gas was bubbled into the **10Fe-MnO$_x$**-TMB system, indicating the crucial role of dissolved oxygen on oxidase activity (Fig. S12). The possible mechanism for oxidase-like activity shown by **10Fe-MnO$_x$** is illustrated in Scheme 1.

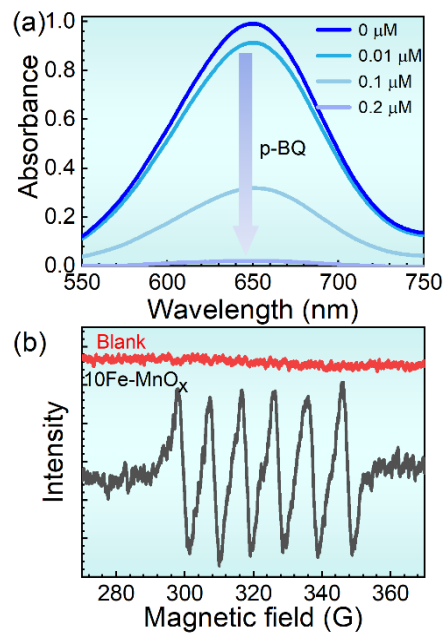

**Fig. 7** (a) Variation in the absorbance of the **10Fe-MnO$_x$**-TMB system upon adding *para*-benzoquinone (*p*-BQ) and (b) EPR spectra of ethanolic solution with and without **10Fe-MnO$_x$**.

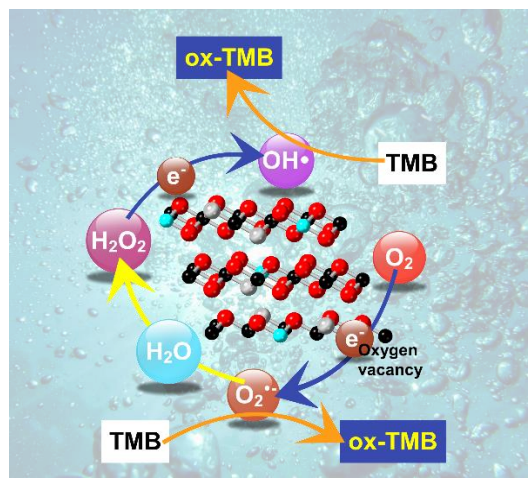

**Scheme 1** Possible mechanism for oxidase-like activity shown by **10Fe-MnO$_x$**. Schematic representation of the crystal structure of **10Fe-MnO$_x$** is shown in the middle. The balls with black, cyan, red and grey colour represent Mn, Fe, O and oxygen vacancy (Ov), respectively.

**Evaluation of antioxidant properties.** Given the strong oxidase-like activity of **10Fe-MnO$_x$**, it can be effectively employed to assess the oxidation resistance and, in particular, the ROS scavenging capabilities of various antioxidants. The antioxidant properties of **10Fe-MnO$_x$** have been evaluated against various antioxidants including Cys, Gua, Dop, Arg, and GSH (Fig. S13a). The addition of these antioxidants to the ox-TMB resulted in varying degrees of color reaction suppression, confirming their antioxidant functionality. Cys and GSH displayed the most significant color as well as spectral changes, likely due to the presence of their sulfhydryl (-SH) groups.[50,51] The color fading trend followed the order of GSH > Cys > Dop > Gua > Arg (Fig.

S13b), indicating that Cys and GSH possess superior ROS scavenging abilities. Furthermore, increasing concentrations of GSH led to a more pronounced decrease in absorbance at 652 nm (Fig. 8a), further validating its antioxidant efficiency. These results demonstrate that the **10Fe-MnO$_x$-TMB** colorimetric system provides a reliable method for antioxidant evaluation (Fig. 8b). Recyclability and reusability are key parameters to determine the reliability and practical utility of a catalytic system. In this effort, the recyclability test has been performed by recovering the catalyst *via* centrifugation after each sensing cycle. Notably, the **10Fe-MnO$_x$** retained its catalytic activity over three cycles, making it a promising and reliable catalytic/sensing platform for further applications (Fig. S14).

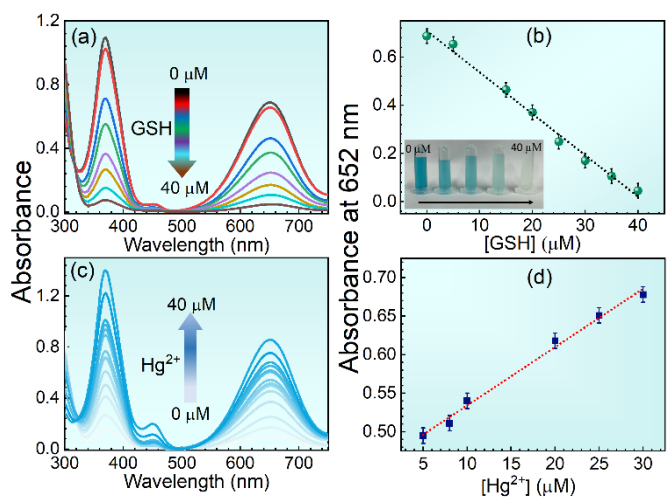

**Fig. 8** (a) The absorption spectra of the 10%Fe-MnO$_x$-TMB system in the presence of GSH with different concentrations (0-40 μM), (b) Calibration plot for the detection of GSH. Inset: photograph of the **10Fe-MnO$_x$-TMB** system with different concentrations of GSH, (c) The absorption spectra of the **10Fe-MnO$_x$-TMB** system in the presence of different concentrations of Hg$^{2+}$ (0-40 μM), (d) Calibration plot for the detection of Hg$^{2+}$.

**Colorimetric detection of Hg$^{2+}$.** Mercury ions are recognized as highly toxic heavy metal pollutants, prevalent in natural environment and a primary contributor to water contamination.[52] Hg$^{2+}$ can be readily transformed into more hazardous organic mercury compounds through microbial activity, leading to bioaccumulation and biomagnification. The Hg$^{2+}$ accumulation in the human body can trigger a range of chronic ailments, including Minamata disease, acrodynia, kidney dysfunction, cognitive impairment, and even damage to the central nervous system.[53-55] Given these health risks, regulatory bodies have established stringent limits for Hg$^{2+}$ in drinking water: the United States Environmental Protection Agency (EPA) sets a maximum contaminant level of 2 μg/mL, while the World Health Organization (WHO) defines a limit of 7 μg/mL. Consequently, the development of sensitive and selective analytical methods for monitoring trace levels of Hg$^{2+}$ in aqueous environments and biological samples is of paramount importance.

As previously discussed, **10Fe-MnO$_x$** demonstrates antioxidant properties in presence of GSH, and the colorimetric signal diminishes significantly due to ROS scavenging ability of -SH group. Under optimized conditions, increasing the concentration of GSH leads to a decrease in absorption intensity at 652 nm (Fig. 8a). Conversely, the addition of varying concentrations of Hg$^{2+}$ to the **10Fe-MnO$_x$/TMB-GSH** system inhibits this decrease in absorbance (Fig. 8c). Therefore, we have employed this system for colorimetric detection of Hg$^{2+}$ (0-40 μM). Notably, **10Fe-MnO$_x$/TMB-GSH** was found to be highly selective for Hg$^{2+}$, with a limit of detection (LOD) depicted as 0.47 μM (Fig. 8d). The colorimetric Hg$^{2+}$ sensing can be attributed to its strong binding capability with glutathione (GSH), ultimately resulting in a concomitant deactivation of antioxidant function. As a result, ROS accumulation restores the colorimetric signal of ox-TMB at 652 nm. The interference study was also conducted in presence of various competing metal ions such as Ag$^+$, Cd$^{2+}$, K$^+$, Na$^+$, Mg$^{2+}$, Ca$^{2+}$, Zn$^{2+}$, Cu$^{2+}$, Co$^{2+}$, Fe$^{2+}$, and Pb$^{2+}$ (Fig. S15) which clearly showed the exceptional selectivity of **10Fe-MnO$_x$/TMB-GSH** towards Hg$^{2+}$.

To further validate the practical applicability of the developed **10Fe-MnO$_x$** system, the colorimetric detection of Hg$^{2+}$ was performed in real water samples spiked with varying concentrations of Hg$^{2+}$. This approach results in excellent recovery rates in the range of 97-105% (Table S3). These findings confirm the reliability of the developed colorimetric assay for Hg$^{2+}$ detection in real water samples.

**Colorimetric detection of hydroquinone.** Hydroquinone, also known as 1,4-dihydroxybenzene, is widely used in various industrial sectors, including cosmetics, dyes, and pesticides. However, its inherent toxicity and limited biodegradability pose significant environmental concerns.[56-58] Excessive exposure to HQ can lead to a range of adverse health effects, such as fatigue, tachycardia, liver dysfunction, and even kidney damage.[59] Therefore, developing accurate, sensitive, rapid, and cost-effective methods for HQ detection is crucial for environmental monitoring and public health protection. Various analytical techniques, including high-performance liquid chromatography (HPLC),[60] capillary electrophoresis,[61] fluorescence spectroscopy,[62] and electrochemical methods,[63] have been well employed for detecting and quantifying HQ. While these techniques offer good sensitivity and specificity, they often suffer from limitations such as time-consuming procedures, reliance on expensive and complex instrumentation, or the need for electrode modification. In contrast, colorimetric methods offer a promising alternative for on-site detection due to their simplicity, speed, and low cost, making them particularly suitable for naked-eye analysis.[64] The selective detection of HQ using a simple colorimetric approach is a meaningful but challenging task in environmental monitoring, with the key requirement being the identification of a suitable chromogenic probe capable of inducing a distinct color change in the presence of HQ.

Here, we employed **10Fe-MnO$_x$**/TMB colorimetric sensing platform for HQ detection. Under optimized conditions, varying concentrations of HQ (0-50 µM) were introduced into the ox-TMB system. As shown in Fig. 9b, the absorbance peak corresponding to ox-TMB at 652 nm decreased with increasing HQ concentration. This enabled the detection of HQ at concentrations as low as 1.74 µM, with a correlation coefficient ($R^2$) of 0.99 (Fig. 9c). The underlying mechanism involves the transfer of two electrons from HQ to ox-TMB due to the strong reducing power of HQ. This process leads to the conversion of HQ to its *p*-benzoquinone tautomer, while the blue ox-TMB is reduced to colorless TMB, as illustrated in Fig. 9a.

To assess the selectivity of the **10Fe-MnO$_x$**/TMB colorimetric system for HQ detection, we tested its response to a range of common interfering substances, including $Na^+$, $K^+$, $Ca^{2+}$, $Mg^{2+}$, $Cu^{2+}$, $Co^{2+}$, $Zn^{2+}$, $Fe^{3+}$, $Pb^{2+}$, $Pd^{2+}$, $Ni^{2+}$ and $Al^{3+}$, phenol, oxalic acid and ethanol. As shown in Fig. S16, these interferents induced only weak absorbance changes at 652 nm compared to the significant decrease observed in the presence of HQ. These results demonstrate that the **10Fe-MnO$_x$**/TMB colorimetric system exhibits high selectivity for HQ detection.

Further to demonstrate the practical utility of the probe in real samples, water samples from different sources were collected and pre-treated. The resulting samples were spiked with different concentration of HQ. The resultant data are summarized in Table S4. The **10Fe-MnO$_x$** based colorimetric method demonstrated accuracy for determining HQ in real samples, with recoveries between 96%-106%.

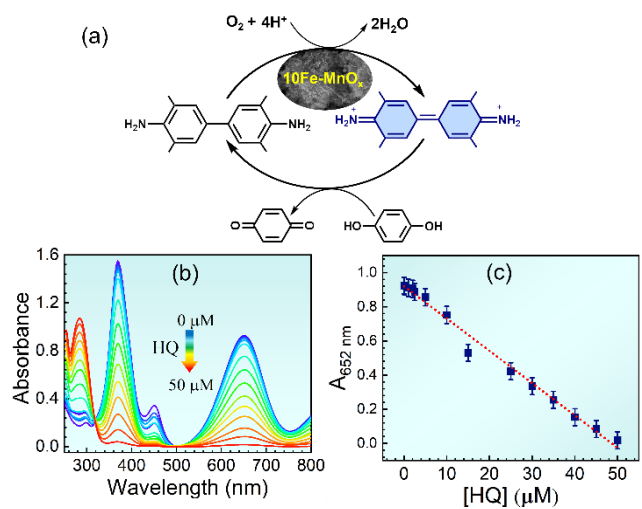

**Fig. 9** (a) Schematic for TMB oxidation and subsequent detection of HQ; (b) UV-Vis spectra of **10Fe-MnO$_x$**/TMB system in the presence of various concentration of HQ (0-50 µM) and (c) Calibration curve for the detection of HQ using **10Fe-MnO$_x$**/TMB.

**Stability and scalability tests of 10Fe-MnO$_x$.** The commercial viability of **10Fe-MnO$_x$** heterogeneous catalyst is significantly dependent on its stability, and scalability. After being stored for one month, the catalyst retained its efficiency, demonstrating good environmental tolerance and surpassing the performance of natural enzymes (Fig. S17a). Furthermore, bulk preparation of the catalyst did not diminish its high efficiency in the oxidation of TMB, highlighting its strong potential for commercial applications (Fig. S17b).

## Conclusions

In summary, we successfully synthesized pure phase of Fe-doped *δ*-MnO$_2$ nanoflowers *via* straightforward chemical oxidation of Fe-doped MnBTC MOF. Systematic Fe-doping into the MOF resulted in phase evolution, structural and morphology modulation which significantly influence oxidase-like activity. Among the synthesized materials, the 10% Fe-doped MnO$_x$ exhibited the highest oxidase-like activity, which could be attributed to a synergistic effect of phase structure, oxygen vacancies, and nanoflower morphology. Under the optimized conditions, **10Fe-MnO$_x$** has been effectively utilized for the colorimetric detection of $Hg^{2+}$ and hydroquinone in aqueous media. The as-developed assay demonstrated remarkable sensitivity, with detection limits of 0.47 µM for $Hg^{2+}$ and 1.74 µM for HQ, outperforming conventional nanozymes. Furthermore, mechanistic insights into the catalytic pathways responsible for oxidation reactions was investigated using steady-state kinetics, scavenging study and EPR spectroscopy.

## Conflicts of interest

"There are no conflicts to declare".

## Acknowledgements

TG, SK and SL thanks UPES for generous support through SEED (UPES/R&D-SoAE/08042024/2) and SHODH (UPES/R&D-SoAE/10042024/8) funding. UD gratefully acknowledged the JRF fellowship from CSIR-India.